\documentclass[11pt,a4paper]{article}
\usepackage[english]{babel}
\usepackage{amsmath,amsthm,amssymb,epsfig,latexsym}
\usepackage{easybmat}
\usepackage{braket} 
\usepackage{color}
\usepackage{ulem}
\usepackage[numbers,square,sort&compress]{natbib}

\newcommand{\bz}{\bar z}

\newcommand{\vk}{\varkappa}
\newcommand{\bu}{\bar u}
\newcommand{\bv}{\bar v}
\newcommand{\eps}{\epsilon}

\newcommand{\bw}{\bar w}

\newcommand{\cN}{\mathcal{N}_n^\nu(\bv)}

\newcommand{\be}[1]{\begin{equation}\label{#1}}
\newcommand{\ba}[1]{\begin{multline}\label{#1}}
\newcommand{\ee}{\end{equation}}
\newcommand{\ea}{\end{eqnarray}}

\newcommand{\tr}{\mathop{\rm tr}}

\newcommand{\rank}{\mathop{\rm rank}}

\newcommand{\dd}{\mathrm{d}}

\newcommand{\CC}{\mathbb{C}}

\newtheorem{prop}{Proposition}[section]

 \makeatletter
 \@addtoreset{equation}{section}
 \makeatother

\newcommand{\bea}{\begin{eqnarray}}
\newcommand{\eea}{\end{eqnarray}}

\begin{document}

\vspace{12pt}

\begin{center}
\begin{LARGE}
{\bf   Scalar products of Bethe vectors in the generalized algebraic Bethe ansatz}%
\end{LARGE}

\vspace{40pt}

{\large G.~Kulkarni\footnote{giridhar.kulkarni@ens-lyon.fr}} \\

Univ Lyon, ENS de Lyon, Univ Claude Bernard Lyon 1, CNRS, Laboratoire de Physique, F-69342 Lyon, France\\

\vspace{40pt}

{\large N.~A.~Slavnov\footnote{nslavnov@mi-ras.ru}}\\
Steklov Mathematical Institute of Russian Academy of Sciences, Moscow, Russia\\

 \vspace{12mm}

\end{center}

\vspace{1cm}


\begin{abstract}
We consider an $XYZ$ spin chain within the framework of the generalized algebraic Bethe ansatz. We study scalar products of the transfer matrix eigenvectors and
arbitrary Bethe vectors. In the particular case of free fermions we obtain explicit expressions for the scalar products with different number of parameters in two Bethe vectors.
\end{abstract}

\vspace{4mm}


\vspace{1cm}

\section{Introduction}

The calculation of scalar products within the framework of the algebraic Bethe ansatz \cite{FadST79,FT79,FadLH96} is an important task. The availability of simple and compact formulas for scalar products of Bethe vectors  allows one to study the form factors and correlation functions of quantum integrable models. Several interesting results in this direction have been obtained in models with rational and trigonometric $R$-matrices
\cite{IzeK84,Kor87,KojKS97,JimMMN92,KitMT00,GohKS04,KitMST05,KitKMST09,KitKMST11,KitKMST12,CauHM05,PerSCHMWA06,PerSCHMWA07,CauCS07,BogIK93L,Sla22L}.

To study the fully anisotropic $XYZ$ Heisenberg chain \cite{Hei28}, the generalized algebraic Bethe ansatz is used \cite{FT79}. This is because the $XYZ$ chain
has an $R$-matrix of an 8-vertex model \cite{Sut70,FanW70,Baxter71,Baxter-book}. As a result, the corresponding monodromy matrix does not have a vacuum vector, which is necessary for constructing the eigenvectors of the transfer matrix in the framework of the traditional algebraic Bethe ansatz.

The generalized algebraic Bethe ansatz allows one to construct eigenvectors of the transfer matrix, as well as to obtain Bethe equations that determine the spectrum of the Hamiltonian. At the same time, the problem of studying the scalar products of generalized Bethe vectors acquires an extremely complex technical character and, until recently, practically has not been researched. Progress was made after the development of a new method based on the reduction of scalar products to a system of linear equations \cite{BS19}. For the generalized algebraic Bethe ansatz, this method allowed us to obtain compact determinant representations for the scalar products of the transfer matrix  eigenvectors (on-shell Bethe vectors) and arbitrary (off-shell) Bethe vectors depending on the same number of parameters \cite{SlaZZ21}. In this paper, we call such scalar products \textit{balanced}.

However, further research in this direction showed that such scalar products are not enough to calculate the form factors of local spin operators. The latter can be reduced to the form factors of the matrix elements of the monodromy matrix using the quantum inverse problem \cite{KitMT99,GohK00,MaiT00}. In turn, actions of the elements of the monodromy matrix on the initial on-shell Bethe vector generate linear combinations of off-shell Bethe vectors, in which the number of parameters may differ by one from the original one \cite{KulS23}. As a result, scalar products arise in which the number of parameters in the left and right vectors do not coincide. This article is devoted to the calculation of such \textit{imbalanced} scalar products.

We use the method of reducing scalar products to a system of linear equations. In principle, this method allows one to obtain the result for the $XYZ$ chain for general values of the coupling constants. However, we are facing technical difficulties that have not yet been overcome. Therefore, in this paper, we consider a special case of the $XYZ$ chain, in which it is equivalent to free fermions (the $XY$ chain)  \cite{LieSM61,McC68,Nie67,KatHS70,PerC77,VaiT78a,Ton81,DlorGS83,IzeKS98}. We hope to generalize the obtained results in the future.

The paper is organized as follows.  In section~\ref{S-N}, we give a brief description of the generalized algebraic Bethe ansatz. Here we introduce a gauge transformation of the monodromy matrix and construct the Bethe vectors. We also introduce here scalar products of Bethe vectors. In section~\ref{S-SP} we derive a cascading system of equations, which relates scalar products with different imbalances. In section~\ref{S-AGTO} we consider  scalar products with imbalance  $\pm2$ (that is, the number of parameters in the left and right vectors differs by $\pm 2$). We prove that these scalar products vanish. This observation allows us
to express scalar products with imbalance $\pm1$  in terms of balanced scalar products in section~\ref{S-SPI1}.

At the end of this paper we have collected basic information about Jacobi theta functions in  appendix~\ref{A-JTF}.
In appendix~\ref{A-ZEV} we discuss non-trivial solutions of special homogenous systems. Finally, in appendix~\ref{A-CIM}, we describe a contour
integral method that allows us to calculate certain sums containing theta functions.

\section{Generalized algebraic Bethe ansatz for the XYZ model\label{S-N}}

In this section, we provide basic information about the description of the $XYZ$ model by the generalized algebraic Bethe ansatz.
The reader can get acquainted with this method in more detail in works \cite{FT79,SlaZZ21}.

\subsection{$R$-matrix and monodromy matrix}

The $XYZ$ spin chain is equivalent to an 8-vertex model.  The corresponding
8-vertex $R$-matrix  has the following form:
\be{R-mat}
R(u)=\begin{pmatrix}
a(u)&0&0&d(u)\\
0&b(u)&c(u)&0\\
0&c(u)&b(u)&0\\
d(u)&0&0&a(u)\\
\end{pmatrix},
\ee
where
\be{abcd}
\begin{aligned}
& a(u)=\frac{2\theta_4(\eta |2\tau ) \, \theta_1(u+\eta |2\tau )\,
\theta_4(u|2\tau )}{\theta_2(0|\tau )\,\theta_4(0|2\tau )},
\\[8pt]
& b(u)=\frac{2\theta_4(\eta |2\tau ) \, \theta_4(u+\eta |2\tau )\,
\theta_1(u|2\tau )}{\theta_2(0|\tau )\,\theta_4(0|2\tau )},
\\[8pt]
& c(u)=\frac{2\theta_1(\eta |2\tau ) \, \theta_4(u+\eta |2\tau )\,
\theta_4(u|2\tau )}{\theta_2(0|\tau )\, \theta_4(0|2\tau )},
\\[8pt]
& d(u)=\frac{2\theta_1(\eta |2\tau ) \, \theta_1(u+\eta |2\tau )\,
\theta_1(u|2\tau )}{\theta_2(0|\tau )\, \theta_4(0|2\tau )}.
\end{aligned}
\ee
The definition of the Jacobi theta functions is given in appendix~\ref{A-JTF}.

To define a monodromy matrix we first introduce a quantum Hilbert space $\mathcal{H}$ and an auxiliary space $\mathcal{H}_0$. The first one is  a tensor
product of local quantum spaces $\mathcal{H}=\mathcal{H}_1\otimes\mathcal{H}_2\otimes\cdots\otimes\mathcal{H}_N$. In its turn, $\mathcal{H}_0\cong\mathbb{C}^2$, and each $\mathcal{H}_k\cong\mathbb{C}^2$.

Then the monodromy matrix of the $XYZ$ chain of the length $N$ is defined as a product of the $R$-matrices acting in $\mathcal{H}_0\otimes\mathcal{H}_k$:
\be{Monod-def}
\mathcal{T}(u)=R_{01}(u-\xi_1)R_{02}(u-\xi_2)\cdots R_{0N}(u-\xi_N),
\ee
where complex parameters $\xi_k$ are called inhomogeneities. In what follows, we will consider only chains with an even number of sites $N$.

The monodromy matrix \eqref{Monod-def} satisfies an
$RTT$-relation
\be{RTT}
R_{12}(u-v)\mathcal{T}_1(u)\mathcal{T}_2(v)=\mathcal{T}_2(v) \mathcal{T}_1(u)R_{12}(u-v),
\ee
which holds in the tensor product $\mathbb{C}^2\otimes\mathbb{C}^2\otimes\mathcal{H}$. The subscripts in \eqref{RTT} show in which of the two auxiliary
spaces $\mathbb{C}^2$ the monodromy matrix $\mathcal{T}_k$ acts nontrivially.  If we write down
the monodromy matrix  as a $2\times2$ matrix in the auxiliary space $\mathcal{H}_0$
\be{Monod-def1}
\mathcal{T}(u)=\begin{pmatrix} A(u)& B(u)\\ C(u)& D(u)
\end{pmatrix},
\ee
then relation \eqref{RTT} defines commutation relations between the
operators $A(u)$, $B(u)$, $C(u)$, and  $D(u)$ acting in $\mathcal{H}$.

The Hamiltonian $H$ of the $XYZ$ chain can be extracted from a  transfer matrix ${\sf T}(u)$. The latter is the trace of the monodromy matrix with respect to the auxiliary space
\be{Transf-mat}
{\sf T}(u)={\tr}_0 \mathcal{T}(u)=A(u)+D(u).
\ee
It is the generating function of the integrals of motion. In particular, let us define $H$ by
\be{Ham-TM}
H=    \frac{2\theta_1(\eta |\tau )}{\theta_1'(0|\tau )}  \frac{\dd}{\dd u}\log {\sf T}(u)\Bigr|_{u=0}
 -\frac{\theta_1'(\eta|\tau )}{\theta'_1(0 |\tau )} N \mathbf{1},
\ee
where $\mathbf{1}$ is the identity operator.  Then in the homogeneous limit $\xi_k=0$, $k=1,\dots,N$, we obtain
\be{HamXYZ}
H=\sum_{j=1}^{N} \Bigl (J_x \sigma^x_j\sigma^x_{j+1}+
J_y \sigma^y_j\sigma^y_{j+1}+J_z \sigma^z_j\sigma^z_{j+1}\Bigr ), \qquad  \sigma^{x,y,z}_{N+1}=\sigma^{x,y,z}_1,
\ee
where spin operators $\sigma^{x,y,z}_k$ are Pauli matrices acting non-trivially in $\mathcal{H}_k$. Numerical coefficients  $J_{x,y,z}$ are given by
$$
J_x=\frac{\theta_4(\eta |\tau )}{\theta_4(0|\tau )}\,, \quad
J_y=\frac{\theta_3(\eta |\tau )}{\theta_3(0|\tau )}\,, \quad
J_z=\frac{\theta_2(\eta |\tau )}{\theta_2(0|\tau )}\,.
$$
They play the role of interaction strength along the axis $x$, $y$, and $z$.

Although only a homogeneous case is needed to construct the Hamiltonian of the $XYZ$ chain, in what follows we will consider a more general
inhomogeneous model \eqref{Monod-def} with arbitrary complex inhomogeneities $\xi_k$. We emphasize, however, that we do this solely for reasons of generality.
In all the formulas below, the homogeneous limit is trivial.

In our calculations, we will mainly focus on the case $\eta=1/2$, which corresponds to $J_z=0$. However, many of the formulas below remain valid for
more general case of rational $\eta$.

\subsection{Gauge transformed monodromy matrix and vacuum vectors}

In models with a 6-vertex $R$-matrix, the eigenvectors of the transfer matrix (on-shell Bethe vectors) are constructed by applying creation operators to the vacuum
vector. There is no such vector in the $XYZ$ model. Therefore, to construct Bethe vectors within the framework of the generalized algebraic Bethe ansatz, we need first to introduce generalized gauge-transformed monodromy matrices
\cite{FT79,SlaZZ21}. Let
\be{GaugeT}
\mathcal{T}_{k,l}(u)=M^{-1}_k(u) \mathcal{T}(u)M_l(u)=
\begin{pmatrix}
A_{k,l}(u)&B_{k,l}(u)
\\
C_{k,l}(u)& D_{k,l}(u)\end{pmatrix}.
\ee
Here
\be{Mku}
M_k(u)=\begin{pmatrix}
\theta_1(s_{k} +u|2\tau )&\gamma_k\theta_1(t_{k} -u|2\tau )
\\
\theta_4(s_{k} +u|2\tau )&\gamma_k\theta_4(t_{k} -u|2\tau )
\end{pmatrix},
\ee
where $s_k=s+k\eta$, $t_k=t+k\eta$,
$s,t\in \CC$ are arbitrary parameters and
\be{gamu}
\gamma_k=\frac2{\theta_2(x_k|\tau)\theta_2(0|\tau)}, \qquad\text{where}\qquad  x_k=x+k\eta, \qquad x=\frac{s+t}2.
\ee
It is easy to check that
\be{detM}
\det M_k(u)= \frac{2\theta_1(y+u|\tau)}{\theta_2(0|\tau)}, \qquad\text{where}\qquad y=\frac{s-t}2.
\ee

For the gauge transformed monodromy matrices, there exists a vector whose properties are similar to those of the vacuum vector in the traditional algebraic Bethe ansatz.
Let us introduce a family of local vacuum vectors $|\omega_k^l\rangle$ parameterized by integer $l$:
\be{vacloc}
|\omega_k^l\rangle=\begin{pmatrix}
\theta_1(s_{k+l-1}+\xi_k|2\tau )
\\
\theta_4(s_{k+l-1}+\xi_k|2\tau )
\end{pmatrix}
 \in \mathcal{H}_k.
\ee
The global vacuum vectors are then defined as
\be{vacglob}
|\Omega ^l\rangle=|\omega_1^l\rangle\otimes |\omega_2^l\rangle\otimes \ldots
\otimes |\omega_N^l\rangle  \quad\in \mathcal{H}.
\ee
Then one can check that
\be{actvac}
\begin{aligned}
& C_{l,l+N}(u)|\Omega^l\rangle=0,
\\
& A_{l,l+N}(u)|\Omega^l\rangle=a(u)|\Omega^{l+1}\rangle,
\\
& D_{l,l+N}(u)|\Omega^l\rangle=d(u) |\Omega^{l-1}\rangle,
\end{aligned}
\ee
where
\be{ad}
a(u)=\prod_{k=1}^N \theta_1 (u - \xi_k  + \eta|\tau),  \qquad d(u)=\prod_{k=1}^N \theta_1 (u - \xi_k|\tau ).
\ee

Similarly, we can define dual (left) local vacuum vectors by
\be{du1}
\langle\bar \omega_k^l|= \Big(-\theta_4(t_{k+l}-\xi_k|2\tau); \theta_1(t_{k+l}-\xi_k|2\tau)\Big)\in\mathcal{H}^*_k\;.
\ee
The global dual (left) vacuum vectors are defined as tensor products of the local ones:
\be{du3}
\langle\bar \Omega^l|=\langle\bar \omega_1^l|\otimes
\langle\bar \omega_2^l|\otimes \ldots \otimes
\langle\bar \omega_N^l|  \quad\in \mathcal{H}^*.
\ee
The action of the operators $A_{l,l+N}(u)$, $D_{l,l+N}(u)$, $B_{l,l+N}(u)$ to the left vacuum is given by
\be{actdu4}
\begin{aligned}
&\langle\bar \Omega^l |B_{l,l+N}(u)=0,\\
&\langle\bar \Omega^l |A_{l,l+N}(u)=
\gamma_l\gamma_{l+N}^{-1}a(u)\langle\bar \Omega^{l-1} |,\\
&\langle\bar \Omega^l |D_{l,l+N}(u)=
\gamma_{l+N}\gamma_{l}^{-1}d(u)\langle\bar \Omega^{l+1} |.
\end{aligned}
\ee
The  Bethe vectors are constructed by the successive action of the operators $B_{k,l}(u)$ on the global vacuum vector.
The dual Bethe vectors are constructed by means of the operators $C_{k,l}(v)$ (see below).

    \subsubsection{Bethe vectors}

Before moving on, we introduce some new notation. Henceforth we will omit the modular parameter in the notation of theta functions
 whenever it is equal to $\tau$, namely, $\theta_a(\cdot)\equiv\theta_a(\cdot|\tau)$.

Let us also introduce two functions that will be often used below
\be{functions}
f(u,v)=\frac{\theta_1(u-v+\eta)}{\theta_1(u-v)},\qquad
h(u,v)=\frac{\theta_1(u-v+\eta)}{\theta_1(\eta)}. 
\ee

In what follows, we will constantly deal with sets of complex variables.
We  denote these sets by a bar: $\bu=\{u_1,\dots,u_m\}$, $\bv=\{v_1,\dots,v_n\}$, etc. As a rule, the number of elements in the
sets is not shown explicitly in the equations, however, we give these cardinalities in
special comments to the formulas. We also introduce special subsets $\bu_j=\bu\setminus\{u_j\}$, $\bu_{j,k}=\bu\setminus\{u_j,u_k\}$ and so on.

In order to make the formulas more compact we use a shorthand notation for products of  functions \eqref{functions}.
 Namely, if the functions  $f$ or $h$ depend on a set (or two sets) of variables, this means that one should take the product over the corresponding set.
For example,
 \be{SH-prodllll}
f(u_j,\bu_j)=\prod_{\substack{u_l\in\bu\\ l\ne j}} f(u_j,u_l),  \quad f(\bv,\bu)=\prod_{\substack{u_l\in\bu\\ v_k\in\bv}} f(v_k,u_l)\qquad\text{etc.}
 \ee
By definition, any product over the empty set is equal to $1$. A double product is equal to $1$ if at least one of the sets
is empty.

We also apply this convention to the products of theta functions, for example,
 \be{SH-prodllllth}
\theta_2(u-\bv)=\prod_{v_k\in\bv} \theta_2(u-v_k),\qquad \theta_1(u_j-\bu_j)=\prod_{\substack{u_l\in\bu\\ l\ne j}} \theta_1(u_j-u_l),   \qquad\text{etc.}
 \ee
In particular, expressions \eqref{ad} for the functions $a(u)$ and $d(u)$ take the form
\be{ad-1}
a(u)=\theta_1 (u - \bar\xi  + \eta),  \qquad d(u)= \theta_1 (u -\bar \xi).
\ee

To construct Bethe vectors  we first introduce a generalized pre-Bethe vector as
\be{psinr}
    |\psi_{n-r}^{l}(\bar u)\rangle=
    B_{l-r-1,l+r+1}(u_{n-r})B_{l-r-2,l+r+2}(u_{n-r-1})\cdots B_{l-n,l+n}(u_1)|\Omega^{l-n}\rangle,
\ee
where $\bu=\{u_1,\dots,u_{n-r}\}$ is a set of arbitrary complex numbers, $n=N/2$, and $r\in\mathbb{Z}$.  A generalized
Bethe vector is then defined as a Fourier transform of \eqref{psinr}. For $\eta=1/2$, it has the following form:
\be{BV2}
|\hat\Psi^\nu_{n-r}(\bu)\rangle =\sum_{l=0}^{3} e^{-\pi i\nu l\eta}|\psi_{n-r}^{l}(\bar u)\rangle, \qquad \nu\in\mathbb{Z}/4\mathbb{Z}.
\ee
If the parameters $\bu$ satisfy a system of Bethe equations (see below), then the generalized
Bethe vector becomes an eigenvector of the
transfer matrix ${\sf T}(u)$ \cite{FT79}. The Bethe vectors are symmetric with respect to the parameters $\bu$ due to the commutation relations between the operators $B_{l-k,l+k}$ and $B_{l-k-1,l+k+1}$ \cite{FT79,SlaZZ21}.

Similarly, we can define generalized dual pre-Bethe vectors
\be{DUb1}
\langle\psi_{n-r}^{l}(\bar v)|=
\langle\bar \Omega ^{l-n}|\bar C_{l-n, l+n}(v_1)\ldots
\bar C_{l-r-2, l+r+2}(v_{n-r-1})\bar C_{l-r-1, l+r+1}(v_{n-r}),
\ee
where $\bar C_{kl}=\gamma_k\gamma_l C_{kl}$, and $\bv=\{v_1,\dots,v_{n-r}\}$ is a set of arbitrary complex numbers.
Then dual Bethe vectors are
\be{dBV2}
\langle\hat\Psi^\nu_{n-r}(\bv)| =\sum_{l=0}^{3} e^{\pi i\nu l\eta}\langle\psi_{n-r}^{l}(\bar v)| \qquad \nu\in\mathbb{Z}/4\mathbb{Z}.
\ee
Just like the vectors \eqref{BV2}, the dual Bethe vectors are symmetric with respect to the parameters $\bv$.

Generically, (dual) Bethe vectors become eigenvectors of the transfer matrix (on-shell Bethe vectors) for $r=0$. Let us introduce
\be{chinu}
\chi_\nu(z)=(-1)^ne^{i\pi\eta\nu}a(z)+e^{-i\pi\eta\nu}d(z).
\ee
Then $|\hat\Psi^\nu_{n}(\bu)\rangle$ is an on-shell Bethe vector provided
\be{BE1}
\chi_\nu(u_j)=0,\qquad j=1,\dots,n.
\ee
A set of equations \eqref{BE1} is nothing but a system of Bethe equations in the  case of free fermions $\eta=1/2$.

Similarly, $\langle\hat\Psi^\nu_{n}(\bv)|$ is a dual eigenvector of the transfer matrix (dual on-shell Bethe vector)  provided
\be{BE2}
\chi_\nu(v_j)=0,\qquad j=1,\dots,n.
\ee
Then
\be{actTvect}
\begin{aligned}
&{\sf T}(z)|\hat\Psi^\nu_n(\bu)\rangle=T_\nu(z|\bu)|\hat\Psi^\nu_n(\bu)\rangle,\\
&\langle\hat\Psi^\nu_{n}(\bv)|{\sf T}(z)=T_\nu(z|\bv)\langle\hat\Psi^\nu_{n}(\bv)|.
\end{aligned}
\ee
where
\be{Tnu1}
T_\nu(z|\bu)=\chi_\nu(z)f(z,\bu),\qquad T_\nu(z|\bv)=\chi_\nu(z)f(z,\bv).
\ee

In the $XYZ$ model with a rational value of $\eta$, there is a degeneracy of the spectrum \cite{FMC03,FabM05,FabM04}. In particular, for $\eta=1/2$, the degeneracy is due to the presence of roots of Bethe equations differing from each other by $1/2$.
Let us define the following mapping over the fundamental domain $z\in\mathbb C/(\mathbb Z+\tau\mathbb Z)$
\begin{equation}
    z^\ast = z+\frac{(-1)^\eps}{2},
\end{equation}
where $\eps=0$ if $0\leq\Re z<\frac12$ and $\eps=1$ otherwise.
It is easy to check that $\chi_\nu(z^\ast)=(-1)^{\nu}\chi_\nu(z)$. Therefore, if $v_a$ is a root $\chi_\nu(z)$, then $v_a^\ast$ is also a root $\chi_\nu(z)$. We will call such elements twins. Since equation \eqref{chinu} is an elliptic polynomial of degree $N$, we conclude that
we have $N/2$ vacancies (i.e. possible solutions) of \eqref{BE2} in the half domain $0\leq \Re z<\frac12$ and $N/2$ vacancies in the remaining half $\frac12\leq \Re z<1$ of the fundamental domain.

In what follows, we will work only with twin-free on-shell Bethe vectors that correspond to singlet eigenstates. Consideration of
vectors with twins requires a special study (see e.g. \cite{FabM06,Deg02,Deg02a,Fab07}.

If a twin-free (dual) on-shell Bethe vector is parameterized by the roots $\bv$, then it follows from \eqref{Tnu1} that the eigenvalues are zero for any of its twin roots:
\begin{equation}
    T_\nu(v_a^\ast)=0,\,\forall a.
\end{equation}
Whereas the eigenvalues of the transfer matrix can be evaluated at its Bethe roots and  written as $T_\nu(v_a)=\theta_2(0)\Omega^\nu_a$ where
\begin{equation}
    \Omega^\nu_a =
    \frac{1}{\theta_2(0)}
    \lim_{z\to v_a}
    T_\nu(z)
    =
    \frac{(-1)^n e^{i\pi\nu\eta} a(v_a) f(v_a,\bv_a) \mathcal V_a}{\theta_1'(0)}
    =
    \frac{- e^{-i\pi\nu\eta} d(v_a) f(v_a,\bv_a) \mathcal V_a}{\theta_1'(0)}.
    \label{omgdef}
\end{equation}
Here we introduced a logarithmic derivative
\begin{equation}
    \mathcal V_a=
    \frac{\dd}{\dd z}\left(\frac{a(z)}{d(z)}\right)\bigg|_{z=v_a},
    \label{Vadef0}
\end{equation}
which appears in normalized expressions for scalar products (see below).

\subsection{Scalar products, selection rule}
In \cite{SlaZZ21}, scalar products of the following form  were considered:
\be{SPdef}
\mathbf S^{\nu,\lambda}_{n,n}(\bv|\bu)=\cN\langle\Psi^\nu_n(\bv)|\Psi^\lambda_n(\bu)\rangle,
\ee
where
\be{Norm}
\cN=\left(\langle\Psi^\nu_n(\bv)|\Psi^\nu_n(\bv)\rangle\right)^{-1}.
\ee
In these equations, $\langle\Psi^\nu_n(\bv)|$ is a dual on-shell Bethe vector, that is the set $\bv$ satisfies Bethe equations \eqref{BE2}. At the same time, $|\Psi^\lambda_n(\bu)\rangle$ is an off-shell Bethe vector, that is, the set $\bu$ does not enjoy any constraint. However, we require that $\#\bv=\#\bu=n$.
We call such scalar products balanced.

As we have shown in \cite{KulS23}, the scalar products \eqref{SPdef} are not enough for the description of the form factors of local operators. Therefore, in this paper, we consider scalar products of a more general form
\be{SPdefn}
\mathbf S^{\nu,\lambda}_{n,m}(\bv|\bu)=\cN\langle\Psi^\nu_n(\bv)|\Psi^\lambda_m(\bu)\rangle.
\ee
As before, $\langle\Psi^\nu_n(\bv)|$ is a dual on-shell Bethe vector. Moreover, we assume that the set $\bv$ does not contain twins. As for the vector $|\Psi^\lambda_m(\bu)\rangle$, it is still an off-shell Bethe vector, however, $\#\bu=m$, and $m$
may be different from $n$. We will call $\vk=n-m$ an imbalance and the corresponding scalar products imbalanced.

It is easy to establish some properties of scalar products with respect to the imbalance. They are based on the properties of the operator ${\sf U}_3$:
\be{UU}
{\sf U}_3=\sigma_1^z\otimes\cdots \otimes\sigma_N^z.
\ee
It was shown in \cite{SlaZZ21} that generic (dual) off-shell Bethe vectors are eigenvectors of ${\sf U}_3$:
\be{nu11}
{\sf U}_3|\Psi^\lambda_m(\bu)\rangle = (-1)^{\lambda+m} |\Psi^\lambda_m(\bu)\rangle,\qquad
\langle\Psi^\nu_n(\bv)|{\sf U}_3 = (-1)^{\nu+n} \langle\Psi^\nu_n(\bv)|.
 \ee
This implies a selection rule
\be{Ortho}
\mathbf S^{\nu,\lambda}_{n,m}(\bv|\bu)\cong \delta_{\nu+\vk,\lambda\;(\hspace{-3mm}\mod 2)}\;.
\ee
In the case of free fermions, it follows from \eqref{Ortho} that either $\lambda=\nu$ or $\lambda=\nu+2\hspace{-1mm}\mod 2$.

Note also that scalar products are symmetric in $\bv$ and symmetric in $\bu$.


\section{Cascading systems of linear equations for scalar products\label{S-SP}}

We use the approach developed in \cite{BS19} that was also used in \cite{SlaZZ21} for the balanced scalar products of the XYZ model. From now on, we consider the free fermion case only, so the parameter $\eta=1/2$ is fixed. However some part of the arguments, including initial remarks, still holds for rational $\eta$ models.
`

Let $\langle\hat\Psi^\nu_n(\bv)|$ be a twin-free dual eigenstate, and let $\bu$ be a set of generic complex numbers of cardinality $\#\bu=n-\vk+1$.
Consider all possible subsets $\bu_j$, $j=1,2,\ldots,\#\bu$, of this set and define corresponding Bethe vectors $|\hat\Psi^\lambda_{m}(\bu_j)\rangle$,
where $m=n-\vk$. Then we can introduce the following imbalanced scalar products:
\be{Xla}
X^{\lambda}_j=\cN\langle\hat\Psi^\nu_n(\bv)|\hat\Psi^\lambda_{m}(\bu_j)\rangle.
\ee
Thus, we get $2(n-\vk+1)$  variables $X^\lambda_j$, taking into account all possible values of $\lambda$ and $j$ and the selection rule \eqref{Ortho}. It is important to stress that, by definition, $X^\lambda_j$ does not depend on $u_j$ and is a symmetric function of $\bu_j$. On the other hand, $X^\lambda_j$ depends on the imbalance, but for the sake of brevity, we do not explicitly specify the parameter $\vk$ in our variable notation.

One can derive a system of linear equations for variables $X^\lambda_j$ using the following procedure.
Inserting a transfer matrix ${\sf T}(u_j)$ between two vectors we obtain on the one hand
\be{Sandwgen-1}
   \cN \langle\hat\Psi^\nu_n(\bv)|{\sf T}(u_j)|\hat\Psi^\lambda_m(\bu_j)\rangle = T_\nu(u_j|\bv)X_j^\lambda,
\ee
where $T_\nu$ is the eigenvalue of transfer matrix for the dual on-shell Bethe vector \eqref{Tnu1}.
On the other hand, we can compute the action of ${\sf T}(u_j)$ to the right using the action formulas \cite{KulS23}.
These formulas were obtained for the $XYZ$ chain with rational $\eta$ of the form $\eta=2P/Q$, where $P$ and $Q$ are coprime integers.
Schematically, the action of the transfer matrix ${\sf T}(u_j)$ on the Bethe vector $|\hat\Psi^\lambda_m(\bu_j)\rangle$
can be written as follows:
\begin{multline}\label{actABCD-BV}
{\sf T}(u_j)|\hat\Psi^\lambda_m(\bu_j)\rangle
=\sum_{\mu=0}^{Q-1}\Bigg\{\sum_{k=1}^{m+1}
\mathbf{W}^{(\lambda-\mu)}_{\vk;1}(u_j,u_k)\chi_\mu(u_k)|\hat\Psi^\mu_m(\bu_k)\rangle \\
%
+\sum_{a>b}^{m+1}
	\mathbf{W}^{(\lambda-\mu)}_{\vk;2}(u_j,u_a,u_b)|\hat\Psi^\mu_{m-1}(\bu_{a,b})\rangle
+\mathbf{W}^{(\lambda-\mu)}_{\vk;3}(u_j)|\hat\Psi^\mu_{m+1}(\bu)\rangle\Bigg\}.
\end{multline}
Here $\mathbf{W}^{(\lambda)}_{\vk;k}$ are some numerical coefficients  (see \cite{KulS23} for their detailed description). We give their explicit form for specific cases below. In particular,
these coefficients depend on the imbalance $\vk$.

In our case $P=1$, $Q=4$. Then
equation \eqref{actABCD-BV} immediately implies
\begin{multline}\label{actABCD-CVBV}
\cN\langle\hat\Psi^\nu_n(\bv)|{\sf T}(u_j)|\hat\Psi^\lambda_m(\bu_j)\rangle
=\sum_{\mu=0}^{3}\Bigg\{\sum_{k=1}^{m+1}
\mathbf{W}^{(\lambda-\mu)}_{\vk;1}(u_j,u_k)\chi_\mu(u_k)X_k^\mu\\
%
+\sum_{a>b}^{m+1}
	\mathbf{W}^{(\lambda-\mu)}_{\vk;2}(u_j,u_a,u_b)\mathbf{S}_{n,m-1}^{\nu,\mu}(\bv|\bu_{a,b})
+\mathbf{W}^{(\lambda-\mu)}_{\vk;3}(u_j)\mathbf{S}_{n,m+1}^{\nu,\mu}(\bv|\bu)\Bigg\}.
\end{multline}
Taking into account \eqref{Sandwgen-1} we arrive at a system of equations
\begin{multline}\label{actABCD-syseq}
T_\nu(u_j|\bv)X_j^\lambda-\sum_{\mu=0}^{3}\sum_{k=1}^{m+1}
\mathbf{W}^{(\lambda-\mu)}_{\vk;1}(u_j,u_k)\chi_\mu(u_k)X_k^\mu\\
%
=\sum_{\mu=0}^{3}\Bigg\{\sum_{a>b}^{m+1}
	\mathbf{W}^{(\lambda-\mu)}_{\vk;2}(u_j,u_a,u_b)\mathbf{S}_{n,m-1}^{\nu,\mu}(\bv|\bu_{a,b})
+\mathbf{W}^{(\lambda-\mu)}_{\vk;3}(u_j)\mathbf{S}_{n,m+1}^{\nu,\mu}(\bv|\bu)\Bigg\}.
\end{multline}

Thus, we have obtained a system of linear equations that relates the scalar products with an imbalance $\vk$ and $\vk\pm1$. In the case of free fermions
coefficients $\mathbf{W}^{(\lambda)}_{\vk;2}$ and $\mathbf{W}^{(\lambda)}_{\vk;3}$ vanish at an even imbalance $\vk=2p$. A closed homogeneous system of equations for scalar products with imbalance $\vk=2p$  arises. For $p=0$ this system was considered and solved in \cite{SlaZZ21} for arbitrary rational $\eta$. We give an explicit solution for $\eta=1/2$ below.

In the next section, we will show that in the case of free fermions and imbalance $\vk=\pm2$, homogeneous systems of equations have only trivial solutions consistent with the requirement that $X_j^\lambda$ does not depend on $u_j$. Hence it follows that scalar products with imbalance $\vk=\pm1$ satisfy inhomogeneous systems of equations in which the inhomogeneous part is expressed through the balanced scalar products. Thus, the solutions of these systems are uniquely determined\footnote{The question of the existence of solutions to inhomogeneous systems is not raised, since scalar products with imbalance $\vk=\pm1$ obviously exist.}.

It is important to note that similar arguments can also be used for more general models with rational values of $\eta$. In this case, we would obtain a homogenous system for scalar products with $\vk= 0\mod M(Q)$, where $M(Q)=Q$ for $Q$ odd and $M(Q)=Q/2$ for $Q$ even.
In all other cases, we obtain an inhomogeneous system where the inhomogeneous part is given in terms of scalar products with $\vk'=\vk\pm 1$, which in their turn can be solutions of an inhomogeneous system. Hence, to determine such scalar products where $\vk\neq 0\mod M(Q)$ we are required to consider scalar products with $\vk'=\vk\pm 1$ and so on until this cascading process is terminated by homogenous case $\vk=0\mod M(Q)$. Therefore, the number of intermediate levels with inhomogeneous terms that we need to consider is $M(Q)-1$. Here, for our model $Q=4$, hence we have only one inhomogeneous system to resolve.

Before moving on, we recall results for balanced scalar products $\vk=0$ obtained in \cite{SlaZZ21}. Here we consider the case $\eta=1/2$ only. Then
%
\begin{equation}
    \mathbf S^{\nu,\mu}_{n,n}(\bv|\bu)=
    \phi_1^{\nu,\mu}(S,x)
    \frac{\theta_1'(0|2\tau)}{\theta_1'(0|\tau)}
    \frac{\prod_{a<b}^n\theta_2(v_a-v_b)\theta_2(u_a-u_b)}{\prod_{a,b=1}^n\theta_2(u_a-v_b)}
    \prod_{k=1}^{n}
    \frac{T_\nu(u_k|\bv)}{\Omega^\nu_k}.
    \label{SPeqn}
\end{equation}
Here
\begin{equation}
    \phi^{\nu,\mu}_1(S,x)=
    \delta_{\nu,\mu \mathop{\mathrm{mod}} 2}
    e^{i\pi(\mu-\nu)x}
    \frac{\theta_1(S)\theta_1(S+2x+\frac{(\mu-\nu)\tau}2|2\tau)}{\theta_1(S+\frac{(\mu-\nu)\tau}2|2\tau)\theta_1(2x|2\tau)},
    \label{phi1def}
\end{equation}
where $S=\sum_{j=1}^n(v_j-u_j)$  and $x=(s+t)/2$ (see \eqref{gamu}). The functions $T_\nu(u_k|\bv)$ and $\Omega^\nu_k$ respectively are given by \eqref{Tnu1} and \eqref{omgdef}.

It is convenient to introduce
\begin{equation}
    \mathbf S_{n,n}^{\nu;\epsilon}(\bv|\bu)=
    \mathbf S_{n,n}^{\nu,\nu}(\bv|\bu)+
    (-1)^\epsilon
    \mathbf S_{n,n}^{\nu,\nu+2}(\bv|\bu).
    \label{SPeqepsdef}
\end{equation}
Then one can show using \eqref{the4-the1}--\eqref{the2-the1u0}  that
\begin{equation}
    \mathbf S_{n,n}^{\nu;\epsilon}(\bv|\bu)=
    \frac{\theta_1(S+x_\epsilon)}{\theta_1(x_\epsilon)}
    \frac{\prod_{a<b}^n\theta_2(v_a-v_b)\theta_2(u_a-u_b)}{\prod_{a,b=1}^n\theta_2(u_a-v_b)}
    \prod_{k=1}^{n}
    \frac{T_\nu(u_k|\bv)}{\Omega^\nu_k}.
    \label{SPeqneps}
\end{equation}
Recall that $x_\epsilon=(s+t)/2+\eta\epsilon$.

\section{Scalar products with imbalance $|\vk|=2$\label{S-AGTO}}

We start our considerations with the scalar products with imbalance $|\vk|=2$. In this section, we prove the following proposition.
\begin{prop}\label{prop-1}
Let $\#\bv=n$ and $\langle\hat\Psi_n^\nu(\bv)|$ be a twin-free on-shell Bethe vector. Let $\ket{\hat\Psi^\lambda_m(\bu)}$ be an arbitrary Bethe vector with $m=\#\bu=n\pm2$. Then
\be{SP-prop}
\langle\hat\Psi^\nu_n(\bv)|\hat\Psi^\lambda_{n\pm 2}(\bu)\rangle=0.
\ee
\end{prop}

The proof consists of several steps. First of all, we show that the scalar products under consideration satisfy  homogeneous systems of linear equations.
Then we transform this system into a new one. This transformation is common for $\vk=2$ and $\vk=-2$. Finally, we prove that the obtained systems have only trivial solutions. The corresponding proofs are different for $\vk=2$ and $\vk=-2$.

\subsection{Homogeneous system of equations}

As we have mentioned above,  the coefficients $\mathbf{W}^{(\lambda-\mu)}_{\vk;2}$ and $\mathbf{W}^{(\lambda-\mu)}_{\vk;3}$ vanish for even values $\vk=2p$. The system \eqref{actABCD-syseq} becomes homogenous.
To describe the coefficients $\mathbf{W}^{(\lambda-\mu)}_{\vk;1}$ we first introduce a function
\begin{equation}
    \alpha_{l}(z)=
    \frac{\theta_2(z+x_l)}{\theta_1(x_l)},
    \label{alp}
\end{equation}
and its Fourier transform
\begin{equation}
    \hat\alpha_\mu(z)=\sum_{l=0}^3e^{-i\pi\eta\mu l}\alpha_l(z), \qquad \mu\in\mathbb{Z}/4\mathbb{Z}.
    \label{alpFT}
\end{equation}
Then
\be{Wvk1}
\mathbf{W}^{(\lambda-\mu)}_{\vk;1}(u_j,u_k)=\frac14\frac{f(u_k,\bu_k)}{h(u_j,u_k)}\hat\alpha_{\lambda-\mu}(u_{jk}),
\ee
and we obtain
\be{Sandw-4}
f(u_j,\bv)\chi_\nu(u_j) X_j^\lambda - \frac14\sum_{\mu=0}^3
\sum_{k=1}^{n-2p+1}\frac{f(u_k,\bu_k)}{h(u_j,u_k)}\hat\alpha_{\lambda-\mu}(u_{jk})\chi_\mu(u_k) X_k^{\mu}=0.
\ee
Here and further $u_{jk}=u_j-u_k$, $v_{nl}=v_n-v_l$, and so on.

It is easy to see that $\hat\alpha_{1}(z)=\hat\alpha_{3}(z)=0$. Besides, the selection rule \eqref{Ortho} imposes that either $\lambda=\nu$ or $\lambda=\nu+2$. Hence, we obtain
\be{2-eq}
\begin{aligned}
&f(u_j,\bv)\chi_\nu(u_j) X_j^\nu = \frac14
\sum_{k=1}^{n-2p+1}\frac{f(u_k,\bu_k)}{h(u_j,u_k)}\chi_\nu(u_k)\Big(\hat\alpha_{0}(u_{jk}) X_k^{\nu}-
\hat\alpha_{2}(u_{jk}) X_k^{\nu+2}\Big),\\
&f(u_j,\bv)\chi_\nu(u_j) X_j^{\nu+2} = \frac14
\sum_{k=1}^{n-2p+1}\frac{f(u_k,\bu_k)}{h(u_j,u_k)}\chi_\nu(u_k)\Big(\hat\alpha_{2}(u_{jk}) X_k^{\nu}-
\hat\alpha_{0}(u_{jk}) X_k^{\nu+2}\Big).
\end{aligned}
\ee

Let
\be{Xeps}
 X_j^\eps= X_j^\nu+(-1)^\eps  X_j^{\nu+2}, \qquad \eps=0,1.
\ee
Then the system \eqref{Sandw-4} can be rewritten as
\begin{equation}\label{hom-sys-1}
\chi_\nu(u_j) X_j^\eps -\frac{1}{f(u_j,\bv)}
\sum_{k=1}^{n-2p+1}\frac{f(u_k,\bu_k)}{h(u_j,u_k)}\chi_\nu(u_k)\alpha_{\eps}(u_{jk}) X_k^{1-\eps} =0, \qquad \eps=0,1,
\end{equation}
where we used $\hat\alpha_{0}+  (-1)^\eps \hat\alpha_{2} =4\alpha_\eps$. Setting
\be{XtX}
 X_j^\eps= Y_j^\eps \prod_{\substack{a=1\\ a\ne j}}^{n-2p+1}\chi_\nu(u_a),
\ee
we exclude the function $\chi_\nu$ from equations \eqref{hom-sys-1}:
\begin{equation}\label{hom-sys-2}
Y_j^\eps -\frac{1}{f(u_j,\bv)}
\sum_{k=1}^{n-2p+1}\frac{f(u_k,\bu_k)}{h(u_j,u_k)}
\alpha_{\eps}(u_{jk})Y_k^{1-\eps} =0, \qquad \eps=0,1.
\end{equation}
It is important to note that $Y_j^\eps$ does not depend on $u_j$ and is a symmetric function of $\bu_j$.
Let us recall that $X_j^\epsilon$ and hence $Y_j^\epsilon$ are also functions of parameters $\bar v$. Although the set  $\bar v$ is fixed by Bethe equations, we can still consider $v_k$ as free parameters, by imposing Bethe equations on inhomogeneities $\bar\xi$ instead of
spectral parameters $\bar v$.

\subsubsection{Transformation of the  system\label{SS-TS}}

The matrix of the system has the following block form:
\be{Msys}
\mathbf{M}=\begin{pmatrix} \mathbf{I}& \mathbf{\Omega}^0  \\ \mathbf{\Omega}^1 &\mathbf{I}
\end{pmatrix},
\ee
where each block is an $(n-2p+1)\times(n-2p+1)$ matrix and
\be{Omell}
\mathbf{\Omega}^\eps_{jk}=-\frac{1}{f(u_j,\bv)}\frac{f(u_k,\bu_k)}{h(u_j,u_k)}\alpha_{\eps}(u_{jk}).
\ee
Setting $\eps=0$ in \eqref{hom-sys-2} we obtain
\be{Y01}
Y_j^0=-\sum_{k=1}^{n-2p+1} \mathbf{\Omega}^0_{jk}Y^1_k.
\ee
Substituting this into  \eqref{hom-sys-2} with $\eps=1$ we arrive at
\be{Y11}
\sum_{k=1}^{n-2p+1}\big(\mathbf{I}- \mathbf{\Omega}^1 \mathbf{\Omega}^0\big)_{jk}Y^1_k=0.
\ee

The product $\mathbf{\Omega}^1\mathbf{\Omega}^0$  can be explicitly calculated using a contour integral method (see appendix~\ref{A-CIM}). We have
\begin{equation}\label{OmOm-3}
\left(\mathbf{I}-\mathbf{\Omega}^1\mathbf{\Omega}^0\right)_{jk}= -\frac{\theta_2^2(0)}{\theta_1(x)\theta_2(x)}
\sum_{q=1}^n{A}_{jq} {B}_{qk},
\end{equation}
where
\be{A-en}
{A}_{jk}= \frac{f(v_k,\bv_k)}{f(u_j,\bv)f(v_k,\bu)}
\frac{\theta_1(u_{j}-v_k+x)}
{\theta_1(u_j-v_k)}, \qquad j=1,\dots,n-2p+1; \quad k=1,\dots,n,
\ee
and
\be{B-en}
{B}_{jk}= \frac{\theta_2(u_{k}-v_j-x)}
{\theta_1(u_k-v_j)}f(u_k,\bu_k),\qquad j=1,\dots,n; \quad k=1,\dots,n-2p+1.
\ee
Thus,
\be{det-1}
\det_{n-2p+1}\left(\mathbf{I}-\mathbf{\Omega}^1\mathbf{\Omega}^0\right)=\left(  -\frac{\theta_2^2(0)}{\theta_1(x)\theta_2(x)} \right)^{n-2p+1}
\det_{n-2p+1}\left(\sum_{q=1}^n{A}_{jq} {B}_{qk}\right).
\ee

\subsection{Scalar products with imbalance $\vk=2$}

The imbalance $\vk=2$ corresponds to $p=1$. We will prove that the determinant \eqref{det-1} is non-vanishing in this case. Note, that this
determinant is an analytic function of $\bv$, $\bu$, and $x$. Hence, to prove that this function is not identically zero,
it is enough to prove that it is non-vanishing for some special values of $\bv$, $\bu$, and $x$. Then, due to analyticity, this function
is non-vanishing in some vicinity of this special point. Hence, it is not identically zero.

\textsl{Remark.} Note that if the set $\bv$ contained twins, then some terms in the sum over $q$ in \eqref{det-1} would vanish.
Indeed,  it follows from \eqref{A-en} that the coefficients $A_{jk}$ are proportional to the products $f(v_k,\bv_k)$. But $f(v_k,\bv_k)$
vanishes if $v_k$ has a twin. On the other hand, if the set $\bv$ does not contain twins, then the sum over $q$ in \eqref{det-1} contains exactly $n$ terms.

Let us introduce $(n-1)\times(n-1)$ matrices $\mathbf{A}$ and $\mathbf{B}$ by
\be{AB}
\mathbf{A}_{jk}={A}_{jk}, \qquad \mathbf{B}_{jk}={B}_{jk}, \qquad j,k=1,\dots,n-1.
\ee
Then
\begin{equation}\label{OmG-3}
\det_{n-1}\left(\sum_{q=1}^n{A}_{jq} {B}_{qk}\right)=\det_{n-1}\left((\mathbf{A}\mathbf{B})_{jk}+{A}_{jn} {B}_{nk}\right)=
\det_{n-1}\mathbf{A}\det_{n-1}\mathbf{B}\det_{n-1}(\mathbf{I}+\mathbf{L}),
\end{equation}
where
\be{bolL}
\mathbf{L}_{jk}=\sum_{\ell,m=1}^{n-1}(\mathbf{A^{-1}})_{j\ell}A_{\ell n}B_{nm}( \mathbf{B^{-1}})_{mk}.
\ee
We see that $\mathbf{L}_{jk}$ is the rank-1 matrix. Hence,
\begin{equation}\label{detG-4}
\det_{n-1}(\mathbf{I}+\mathbf{L})=1+\sum_{k=1}^{n-1}\mathbf{L}_{kk}.
\end{equation}
Both ${\mathbf{A}}$ and ${\mathbf{B}}$ are Cauchy matrices (multiplied with diagonal matrices) with non-vanishing determinants for generic $\bu$, $\bv$, and $x$.
Thus, it is enough to prove that
\be{toProve}
1+\sum_{k=1}^{n-1}\mathbf{L}_{kk}\ne 0,
\ee
for generic $\bu$, $\bv$, and $x$.
%
The inverses of Cauchy matrices $\mathbf A$ and $\mathbf B$ have the following form (see \eqref{BCau1}, \eqref{BCau-inv}):
\be{A-inv}
(\mathbf{A}^{-1})_{jk}=\frac{-f(u_k,\bv)}{f(v_n,v_j)\theta_1(x)\theta_1(x-S)}\frac{\theta_2(\bu-v_j)}{\theta_2(\bv_{n,j}- v_j)}
\frac{\theta_1(u_k-v_j-x+S)}{\theta_1(u_k-v_j)}
\frac{\theta_1(u_k-\bv_n)}{\theta_1(u_k-\bu_k)},
\ee
and
\be{B-inv}
(\mathbf{B}^{-1})_{jk}=\frac{1}{\theta_2(x)\theta_2(x+S)}\frac{\theta_1(\bu-v_k)}{\theta_1(\bv_{n,k}- v_k)}
\frac{\theta_2(u_j-v_k+x+S)}{\theta_1(u_j-v_k)}
\frac{\theta_1(u_j-\bv_n)}{\theta_2(u_j-\bu_j)},
\ee
where
\be{SSS}
S=\sum_{j=1}^{n-1}(v_j-u_j).
\ee
Recall that notation $\theta_2(\bu-v_j)$ and similar ones means the product of theta functions over the set $\bu$ according to convention \eqref{SH-prodllllth}.

Let
\be{calAB}
\begin{aligned}
&\mathcal{A}_j=\sum_{\ell=1}^{n-1}(\mathbf{A^{-1}})_{j\ell}A_{\ell n},\\
&\mathcal{B}_k=\sum_{\ell=1}^{n-1}B_{n\ell}( \mathbf{B^{-1}})_{\ell k}.
\end{aligned}
\ee
Then using the contour integral method we find
\be{calAj-res}
\mathcal{A}_j=\frac{1}{\theta_1(x-S)}\frac{\theta_2(\bu-v_j)}{\theta_2(\bv_{n,j}- v_j)}
\frac{\theta_1(v_{nj}-x-S)}{\theta_2(v_{nj})}
\frac{\theta_2(v_n-\bv_n)}{\theta_2(v_n-\bu)},
\ee
and
\be{calBk-res}
\mathcal{B}_k=\frac{-1}{\theta_2(x+S)}\frac{\theta_1(\bu-v_k)}{\theta_1(\bv_{n,k}- v_k)}
\frac{\theta_2(v_{nk}+x-S)}{\theta_1(v_{nk})}
\frac{\theta_1(v_n-\bv_n)}{\theta_1(v_n-\bu)}.
\ee

Thus, using \eqref{the2-the1} and \eqref{the2-the1uu} we obtain
\begin{multline}\label{calABk-res}
1+\sum_{k=1}^{n-1} \mathbf{L}_{kk}=1+\frac{\theta_1(2v_n-2\bv_n|2\tau)}{\theta_1(2v_n-2\bu|2\tau)}
\\
\times\sum_{k=1}^{n-1}\frac{\theta_1(2\bu-2v_k|2\tau)}{\theta_1(2\bv_{k}-2v _k|2\tau)}
\frac{\theta_1(2x|2\tau)\theta_4(2v_{nk}+2S|2\tau)-\theta_4(2x|2\tau)\theta_1(2v_{nk}+2S|2\tau)}
{\theta_1(2x|2\tau)\theta_4(2S|2\tau)-\theta_4(2x|2\tau)\theta_1(2S|2\tau)}.
\end{multline}

Consider a special case $u_k=v_k$ for $k=2,\dots,n-1$. Then $S=v_1-u_1$. We also see that only the term with $k=1$ survives in the sum \eqref{calABk-res}
due to the product $\theta_1(2\bu-2v_k|2\tau)$. Hence,
\begin{multline}\label{Lkk-1}
1+\sum_{k=1}^{n-1} \mathbf{L}_{kk}=1+\frac{\theta_1(2v_1-2u_1|2\tau)}{\theta_1(2v_n-2u_1|2\tau)}\\
\times
\left(\frac{\theta_1(2x|2\tau)\theta_4(2v_{n}-2u_1|2\tau)-\theta_4(2x|2\tau)\theta_1(2v_{n}-2u_1|2\tau)}
{\theta_4(2x|2\tau)\theta_1(2v_1-2u_1|2\tau)-\theta_1(2x|2\tau)\theta_4(2v_1-2u_1|2\tau)}\right).
\end{multline}
Using identities \eqref{the4-the1} we transform this result as follows:
\be{Lkk-2}
1+\sum_{k=1}^{n-1} \mathbf{L}_{kk}
=\frac{\theta_1(v_1-v_n)\theta_2(v_1+v_n-2u_1)\theta_1(x)\theta_2(x)}
{\theta_1(v_n-u_1)\theta_2(v_n-u_1)\theta_1(v_1-u_1-x)\theta_2(v_1-u_1+x)}.
\ee
We see that $1+\sum_{k=1}^{n-1} \mathbf{L}_{kk}\ne 0$ for generic $v_1$, $v_n$, $u_1$ and $x$.  Thus, we conclude that
$1+\sum_{k=1}^{n-1} \mathbf{L}_{kk}\ne 0$ for generic $\bu$, $\bv$, and $x$. Therefore, the system of equations \eqref{Y11}, and hence system \eqref{hom-sys-2} have only a trivial solution
for $p=1$, and hence,
\be{SP-conj-pr}
\langle\hat\Psi^\nu_n(\bv)|\hat\Psi^\lambda_m(\bu)\rangle=0,
\ee
for $\vk=n-m=2$. This proves part of proposition~\ref{SP-prop}.

\subsection{Scalar products with imbalance $\vk=-2$}

The imbalance $\vk=-2$ corresponds to $p=-1$.
Then the matrix $\mathbf{I}-\mathbf{\Omega}^1\mathbf{\Omega}^0$ has the size $(n+3)\times(n+3)$, and we have
\be{det1-0}
\det_{n+3}\left(\sum_{q=1}^n{A}_{jq} {B}_{qk}\right)=0,
\ee
because the rank of this matrix does not exceed $n$. Hence, in this case, the system \eqref{Y11} has non-trivial solutions. However, we will show that these solutions are not compatible with the condition that $Y^\eps_j$ does not depend on $u_j$.

It is shown in appendix~\ref{A-ZEV} that solutions to the system \eqref{Y11} are expressed in terms of the entries of the inverse matrix $B^{-1}$.
We first should extend the matrix $B_{jk}$ \eqref{B-en} to the size $(n+3)\times(n+3)$. For this, we introduce an extended matrix $\tilde B_{jk}$ by
\be{B-enext}
\tilde {B}_{n+l,k}= \frac{\theta_2(u_{k}-z_l-x)}
{\theta_1(u_k-z_l)}f(u_k,\bu_k),\qquad l\in\{1,2,3\},
\ee
and $\tilde B_{jk}= B_{jk}$ for $j\le n$. Here $z_{1}$, $z_{2}$, and $z_{3}$ are generic complex numbers. Then
\begin{multline}\label{Bext-inv}
(\tilde B^{-1})_{j, n+l}=\frac{1}{\theta_2(x)\theta_2(x+\tilde S)}
\frac{\theta_2(u_j-u_{n+l}+x+\tilde S)}{\theta_1(u_j-u_{n+l})}\\
\times
\frac{\theta_1(u_j-\bv)\theta_1(u_j-\bar z)\theta_1(\bu-z_l)}
{\theta_2(u_j-\bu_j)\theta_1(\bv- z_l)\theta_1(\bar z_l- z_l)},\qquad l=1,2,3,
\end{multline}
where
\be{this-S}
\tilde S=\sum_{j=1}^n(v_j-u_j)+\sum_{l=1}^{3}(z_l-u_{n+l}).
\ee

Thus, we find
\be{solY-0}
Y^1_j=\sum_{l=1}^3 C_lY^{1}_{j;l},
\ee
where
\be{ZevCa33}
Y^{1}_{j;l}=\theta_2(x+u_j-z_{l}-\tilde S)
\frac{\theta_1(u_j-\bar z_{l})\theta_1(u_j-\bar v)}{\theta_2(u_j- \bu_j)},
\ee
and $C_l$ are some functions that depend on the parameters $\bu$ in such a way that $C_lY^{1}_{j;l}$ does not depend on $u_j$.

Substituting $Y^{1}_{j;l}$ \eqref{ZevCa33} into equation \eqref{hom-sys-2} with $\eps=0$ we obtain $Y^{0}_{j;l}$:
\begin{equation}\label{hom-sys-3}
Y_{j;l}^0 =-\frac{C_l\theta_2(0)}{f(u_j,\bv)}
\sum_{k=1}^{n+3}\frac{\theta_1(u_k-\bar z)\theta_1(u_k-\bar v)}{\theta_1(u_k- \bu_k)}
\frac{\theta_1(u_{k}-u_j-x)}{\theta_2(u_k-u_j)\theta_1(x)}
\frac{\theta_2(x+u_k-z_{l}-\tilde S)}{\theta_1(u_k-z_{l})}.
\end{equation}
Computing this sum via the contour integral method we find
\begin{equation}\label{hom-sys-3a}
Y_{j;l}^0 =f(u_j,\bar z_{l})\frac{\theta_2(x)}{\theta_1(x)}\;C_l
\frac{\theta_1(u_j-\bar z_{l})\theta_1(u_j-\bar v)}{\theta_2(u_j- \bu_j)}
\theta_1(x+u_j-z_l-\tilde S).
\end{equation}

We see now that we can not satisfy the condition that $Y^\eps_j$ does not depend on $u_j$. Indeed, to provide this condition for $Y_{j;l}^1$, we
should choose
\be{Ca}
C_l=\tilde C_l\frac{\prod_{a>b}^{n+3}\theta_2(u_a-u_b)}{\theta_1(\bu-\bv)\theta_1(\bu-\bz_{l})},
\ee
where $\tilde C_l$ does not depend on $\bu$. However, it follows form \eqref{hom-sys-3a} that $Y_{j;l}^0$ becomes proportional to the product
$f(u_j,\bar z_{l})$ in this case. And vice versa: choosing the function $C_l$ in \eqref{hom-sys-3a}  to eliminate the dependence on
$u_j$, we thereby create a dependence on $u_j$ in \eqref{ZevCa33}.

Thus, the only solution to system \eqref{hom-sys-2} that satisfies the necessary condition is the trivial solution.
Hence,
\be{SP-conj-pr0}
\langle\hat\Psi^\nu_n(\bv)|\hat\Psi^\lambda_m(\bu)\rangle=0,
\ee
for $\vk=n-m=-2$. Thus, proposition~\ref{SP-prop} is proved.

Through proposition~\ref{prop-1} we have obtained a strong selection rule for the scalar products of twin-free on-shell Bethe vectors. We find that they are orthogonal to all Bethe vectors from the sector $\vk=\pm 2$. However, this does not tell anything about their scalar products with Bethe vectors for $\vk=\pm 1$. In this case, the system of equations \eqref{actABCD-syseq} is inhomogeneous. The next section is devoted to this system.

\section{Imbalanced scalar products with $\vk=\pm 1$\label{S-SPI1}}

It is convenient to make a change of variables in this section. Namely, we replace the set $\bu=\{u_1,\dots,u_{m+1}\}$ with a set $\bw=\{w_1,\dots,w_{m+1}\}$.

For odd values of parameter $\vk=2p+1$, all the coefficients $\mathbf{W}^{(\lambda)}_{\vk;k}$, $k=1,2,3$, survive in the system \eqref{actABCD-syseq}.
To describe them, we first introduce coefficients
\begin{equation}
    \omega_{ab}(z)=
    \frac{\left[d(w_a)a(w_b)-d(w_b)a(w_a)\right]f(w_a,\bw_a)f(\bw_b,w_b)}{f(w_a,w_b)h(w_a,z)h(z,w_b)},
    \label{omg_dbl}
\end{equation}
where $a(u)$ and $d(u)$ are given by \eqref{ad-1}.
Let us also define two functions
\begin{align}
    \beta^+_{l}(z)&=
    \theta_2(z+s_l),
    \label{beta+}
    \\
    \beta^-_{l}(z;u,v)&=
    \theta_2(z-t_l)\theta_2(z-u+x_l)\theta_2(z-v+x_l),
    \label{beta-}
\end{align}
and their Fourier transforms
\begin{equation}
    \hat\beta^\pm_\mu(\cdot)=
    \sum_{l=0}^{3}
    e^{-i\pi\mu\eta l}    \beta^\pm_l(\cdot),  \qquad \mu\in\mathbb{Z}/4\mathbb{Z}.
    \label{betaFT}
\end{equation}
Recall that $s$ and $t$ are the gauge parameters (see \eqref{Mku}), and $s_l=s+l\eta$, $t_l=t+l\eta$.

The system \eqref{actABCD-syseq} then takes the following form
\begin{multline}
    T_\nu(w_j|\bar v)X^\lambda_j
    -\frac{\theta_2(y+w_j)}{4\theta_1(y+w_j)}
    \sum_{\mu=0}^{3}
    \sum_{k=1}^{n-2p}
    \frac{f(w_k,\bw_k)}{h(w_j,w_k)}
    \hat\alpha_{\lambda-\mu}(w_{jk})
    \chi_\mu(w_k)
    X^\mu_k
    \\
    =
    \frac{(-1)^p}{2\theta_1(y+w_j)\theta_1^2(x)\theta_2^2(x)}
    \sum_{\mu=0}^{3}
    \sum_{a>b}^{n-2p}
    \omega_{ab}(w_j)
    \hat\beta^-_{\lambda-\mu}(w_j;w_a,w_b)
    \mathbf{S}_{n,n-2p-2}^{\nu,\mu}(\bar v|\bw_{a,b})
    \\
    +\frac{(-1)^p\theta_2^2(0)}{8\theta_1(y+w_j)}
    \sum_{\mu=0}^{3}
    \hat\beta^+_{\lambda-\mu}(w_j)
    \mathbf{S}_{n,n-2p}^{\nu,\mu}(\bar v|\bw).
    \label{Sinhom-1}
\end{multline}
We see that solutions to inhomogeneous systems for $\vk=\pm 1$ can be found in terms of scalar products with $\vk\in\{-2,0,2\}$. As we have shown in the previous section, these scalar products vanish for $\vk=\pm 2$. At the same time, the case $\vk=0$ is described by \eqref{SPeqneps}. Thus, the inhomogeneous part of the system \eqref{Sinhom-1} is completely determined.

In its turn, the homogeneous part of \eqref{Sinhom-1} can be drastically simplified. Note that we do not need to find $X_j^\lambda$ for all $j=1,\dots,m+1$. It
is enough to find only one of them, for instance, $X_{m+1}^\lambda$. Since $X_{m+1}^\lambda$ does not depend on $w_{m+1}$, we can set
\begin{equation}
     w_{m+1}=-y^\ast=-y+\frac 12.
\end{equation}
Then $\theta_2(y+w_{m+1})=0$, and we immediately obtain an explicit expression for $X_{m+1}^\lambda$:
\begin{multline}\label{Dir-expr}
    X_{m+1}^\lambda=
    \frac{(-1)^{\frac{n-m-1}2}
    }{T_{\nu}(-y^\ast|\bv)}
    \Bigg\lbrace
    \frac{\theta_2(0)}{8}
    \sum_{\mu=0}^{3}
    \hat\beta^+_{\lambda-\mu}(-y^\ast)
    \mathbf S^{\nu,\mu}_{n,m+1}(\bv|\bw)\\
    +\frac{1}{2\theta_1^2(x)\theta_2^2(x)\theta_2(0)}
    \sum_{\mu=0}^{3}
    \sum_{a>b}^{m+1}
    \omega_{ab}(-y^\ast)
    \hat\beta^-_{\lambda-\mu}(-y^\ast;w_a,w_b)
    \mathbf S^{\nu,\mu}_{n,m-1}(\bv|\bw_{a,b})
    \Bigg\rbrace .
\end{multline}
Obviously, $\beta^\pm_{l+2}=-\beta^\pm_l$.
Then it is easy to see that $\hat\beta^\pm_{\lambda}$ is non-zero only for $\lambda=1$ or $\lambda=3$. This implies that $X_{m+1}^{\lambda}$ is zero for $\lambda=\nu$ and $\lambda=\nu+2$ as we expected from the selection rule \eqref{Ortho}. We also see that by virtue of proposition~\ref{prop-1}, the right hand side of \eqref{Dir-expr} contains only one type of scalar product, either
with $m=n-1$ or with $m=n+1$. We consider these two cases separately.

\subsubsection{Imbalanced scalar products with $\vk=1$}

Let $\bw=\{\bu, -y^\ast\}$. Then
\be{X-S}
X^\lambda_{m+1}=\mathbf S^{\nu,\lambda}_{n,n-1}(\bv|\bu).
\ee

Since all the scalar products $\mathbf{S}_{n,n-2}^{\nu,\mu}$ vanish due to proposition~\ref{prop-1}, equation \eqref{Dir-expr} turns into
\begin{align}
    \mathbf S^{\nu,\nu+1}_{n,n-1}(\bv|\bu)&=
    \frac{\theta_2(0)
    }{8T_\nu(-y^\ast|\bv)
    }
    \left[
    \hat\beta^+_{1}(-y^\ast)
    \mathbf S^{\nu,\nu}_{n,n}(\bv|\{\bu,-y^\ast\})
    +
    \hat\beta^+_{3}(-y^\ast)
    \mathbf S^{\nu,\nu+2}_{n,n}(\bv|\{\bu,-y^\ast\})
    \right],
    \\
    \mathbf S^{\nu,\nu+3}_{n,n-1}(\bv|\bu)&=
    \frac{\theta_2(0)
    }{8T_\nu(-y^\ast|\bv)
    }
    \left[
    \hat\beta^+_{1}(-y^\ast)
    \mathbf S^{\nu,\nu+2}_{n,n}(\bv|\{\bu,-y^\ast\})
    +
    \hat\beta^+_{3}(-y^\ast)
    \mathbf S^{\nu,\nu}_{n,n}(\bv|\{\bu,-y^\ast\})
    \right].
\end{align}
Let us introduce
\begin{equation}
    \mathbf S^{\nu;\epsilon}_{n,n-1}(\bv|\bu)
    =
    \mathbf S^{\nu,\nu+1}_{n,n-1}(\bv|\bu)+(-1)^\epsilon
    \mathbf S^{\nu,\nu+3}_{n,n-1}(\bv|\bu), \qquad \epsilon=0,1.
    \label{SPn-1epsdef}
\end{equation}
Then using $\hat\beta_1+ (-1)^\epsilon\hat\beta_3=4(-i)^\epsilon\beta_\epsilon$ we obtain
\begin{equation}
    \mathbf S^{\nu;\epsilon}_{n,n-1}(\bv|\bu)
    =
    -(-i)^\epsilon
    \frac{\theta_2(0)\theta_1(x_\epsilon)}{2T_\nu(-y^\ast|\bv)}
    \mathbf S^{\nu;\epsilon}_{n,n}(\bv|\{\bu,-y^\ast\}),
    \label{SPn-1eps}
\end{equation}
where the scalar product $S_{n,n}^{\nu;\epsilon}(\bv|\bu,-y^\ast)$ on the right hand side is given by \eqref{SPeqneps}.
Replacing in \eqref{SPeqneps} $\bu$ with $\{\bu,-y^\ast\}$, we find
\begin{multline}
    \mathbf S^{\nu;\epsilon}_{n,n-1}(\bv|\bu)=\frac{(-i)^\epsilon}2 \theta_2(0)
    \theta_2(S'+s_\epsilon) \frac{\prod_{a=1}^{n-1} \theta_1(u_a+y)}{\prod_{a=1}^{n}\theta_1(v_a+y)}\\
    \times\frac{\prod_{a<b}^n\theta_2(v_a-v_b)\prod_{a<b}^{n-1}\theta_2(u_a-u_b)}{\prod_{a=1}^{n-1}\prod_{b=1}^n\theta_2(u_a-v_b)}
        \frac{\prod_{k=1}^{n-1} T_\nu(u_k|\bv)}{\prod_{k=1}^{n}\Omega^\nu_k},
    \label{SPimbal-pl1}
\end{multline}
where $S'=\sum_{j=1}^n v_j-\sum_{j=1}^{n-1} u_j$.

\subsubsection{Imbalanced scalar products with $\vk=-1$}

We set again $\bw=\{\bu, -y^\ast\}$ and obtain
\be{X-S1}
X^\lambda_{m+1}=\mathbf S^{\nu,\lambda}_{n,n+1}(\bv|\bu).
\ee
This time, the set $\bu$ consists of $n+1$ parameters.

Due to proposition~\ref{prop-1} all the scalar products $\mathbf{S}_{n,n+2}^{\nu,\lambda}$ are vanishing.
Let us consider a combination similar to \eqref{SPn-1epsdef}
\begin{equation}
    \mathbf S_{n,n+1}^{\nu;\eps}(\bv|\bu)=
    \mathbf S_{n,n+1}^{\nu,\nu+1}(\bv|\bu)
    +(-1)^{\eps}
    \mathbf S_{n,n+1}^{\nu,\nu+3}(\bv|\bu).
    \label{SPn+1epsdef}
\end{equation}
Then similarly to the case $\vk=1$, we obtain
\begin{equation}
    \mathbf S^{\nu;\epsilon}_{n,n+1}(\bv|\bu)=
   \frac{ -2(-i)^\epsilon\theta_1(x_\epsilon)}{\theta_1^2(x)\theta_2^2(x)\theta_2(0)T_\nu(-y^\ast|\bv)}
    \sum_{a>b}^{n+2}\omega_{ab}(-y^\ast)
    \theta_1(w_a-t_\epsilon)
    \theta_1(w_b-t_\epsilon)
    \mathbf{S}^{\nu;\epsilon}_{n,n}(\bv|\bw_{a,b}).
    \label{SPn+1eps_dprod}
\end{equation}
Thus, the scalar product $\mathbf S^{\nu;\epsilon}_{n,n+1}$ is expressed in terms of a simple linear combination of $\mathbf S^{\nu;\epsilon}_{n,n}$.

\section*{Conclusion}

In \cite{KulS23}, we argued that balanced scalar products are not enough to calculate form factors. One also needs to know
scalar products with imbalance $\vk=\pm1$. In this paper, we have shown that in the $XY$ model, the latter are expressed in a simple way through
the balanced scalar products. Thus, we have prepared all the necessary tools for calculating the form factors of local
spin operators in the $XY$ model. Indeed, due to the explicit solution of the inverse problem, the latter are reduced to the form factors  of the
monodromy matrix elements. For example, the form factor of the operator $\sigma_k^z$ is reduced to the matrix element of $A(\xi_k)-D(\xi_k)$:
\be{sigmaz}
\frac{\langle \Psi^\nu_n(\bv)|\sigma_k^z |\Psi^\lambda_n(\bu)\rangle}{\|\Psi^\nu_n(\bv)\| \|\Psi^\lambda_n(\bu)\|}
=
\left(\frac{\prod_{j=1}^{k-1}T_\lambda(\xi_j|\bu)}{\prod_{j=1}^{k}T_\nu(\xi_j|\bv)}\right)
\frac{\langle \Psi^\nu_n(\bv)|\big(A(\xi_k)-D(\xi_k)\big) |\Psi^\lambda_n(\bu)\rangle}{\|\Psi^\nu_n(\bv)\| \|\Psi^\lambda_n(\bu)\|}.
\ee
 The action of $A(\xi_k)-D(\xi_k)$ on the vector $|\Psi^\lambda_n(\bu)\rangle$ is schematically given by formula \eqref{actABCD-BV}
(with other coefficients $\mathbf{W}^{(\lambda)}_{\vk;k}$). Thus, the form factor \eqref{sigmaz} is reduced to a linear combination of scalar products with imbalance $\vk=0,\pm1$.
As for the correct normalization, it is easy to see that it is restored in expressions that are quadratic in form factors.

We believe that the method proposed in this paper can also be generalized to the case of an arbitrary rational value of the parameter $\eta$. As already noted, the cascade system of equations for scalar products \eqref{actABCD-syseq} gives homogeneous equations with an imbalance $\vk= 0\mod M(Q)$, where $M(Q)=Q$ for $Q$ odd and $M(Q)=Q/2$ for Q even. It is most probable that these inhomogeneous equations have only trivial solutions for the same reason as in the case of free fermions. If so, then we get a closed system of inhomogeneous linear equations for scalar products with nonzero imbalance. It remains only to write down the solution of this system in a form convenient for further applications. We will study this issue in our future publications.

\section*{Acknowledgements}
We are grateful to A.~Zabrodin and A.~Zotov for numerous and fruitful discussions. The work of G.K. was supported by the SIMC postdoctoral grant of the Steklov
Mathematical Institute. Section~\ref{S-AGTO} of the paper was performed by N.S.  The work of N.S. was supported by the Russian Science Foundation under grant
no.19-11-00062, https://rscf.ru/en/project/19-11-00062/ , and performed at Steklov Mathematical Institute of Russian Academy of Sciences.

\appendix

\section{Jacobi theta functions\label{A-JTF}}
Here we only give some basic properties of Jacobi theta functions used in the paper. See \cite{KZ15} for more details.

The Jacobi theta functions are defined as follows:
\be{JacTF}
\begin{aligned}
&\theta_1(u|\tau )=-i\sum_{k\in \mathbb{Z}}
(-1)^k q^{(k+\frac{1}{2})^2}e^{\pi i (2k+1)u},
\\[6pt]
&\theta_2(u|\tau )=\sum_{k\in \mathbb{Z}}
q^{(k+\frac{1}{2})^2}e^{\pi i (2k+1)u},
\\[6pt]
&\theta_3(u|\tau )=\sum_{k\in \mathbb{Z}}
q^{k^2}e^{2\pi i ku},
\\[6pt]
&\theta_4(u|\tau )=\sum_{k\in \mathbb{Z}}
(-1)^kq^{k^2}e^{2\pi i ku},
\end{aligned}
\ee
where $\tau \in \mathbb{C}$, $\Im \tau >0$, and
$q=e^{\pi i \tau}$.

They satisfy the following shift properties:
\be{A-shift}
\begin{aligned}
&\theta_1(u+1/2|\tau)=\theta_2(u|\tau),&\qquad &\theta_2(u+1/2|\tau)=-\theta_1(u|\tau),\\
&\theta_1(u+1|\tau)=-\theta_1(u|\tau),&\qquad &\theta_2(u+1|\tau)=-\theta_2(u|\tau),\\
&\theta_1(u+\tau|\tau)=-e^{-\pi i(2u+\tau)}\theta_1(u|\tau),&\qquad &\theta_2(u+\tau|\tau)=e^{-\pi i(2u+\tau)}\theta_2(u|\tau).
\end{aligned}
\ee
In this paper, we use an identity
\be{the4-the1}
2\theta_1(u+v|2\tau)\theta_4(u-v|2\tau)=\theta_1(u|\tau)\theta_2(v|\tau)+\theta_2(u|\tau)\theta_1(v|\tau).
\ee
It follows from this identity that
\be{the2-the1}
\theta_1(u|\tau)\theta_2(v|\tau)=\theta_1(u+v|2\tau)\theta_4(u-v|2\tau)+\theta_4(u+v|2\tau)\theta_1(u-v|2\tau).
\ee
In particular, setting $v=u$ in \eqref{the2-the1} we obtain
\be{the2-the1uu}
\theta_1(u|\tau)\theta_2(u|\tau)=\theta_1(2u|2\tau)\theta_4(0|2\tau).
\ee
Differentiating \eqref{the2-the1uu} with respect to $u$ at zero, we arrive at
\be{the2-the1u0}
\frac{\theta'_1(0|\tau)}{\theta'_1(0|2\tau)}=2\frac{\theta_4(0|2\tau)}{\theta_2(0|\tau)}.
\ee
%

\section{Zero eigenvectors\label{A-ZEV}}

Let $M$ be an $n\times n$ matrix. Let its matrix elements can be presented in the form
\be{formM}
M_{jk}=\sum_{\ell=1}^m A_{j\ell}B_{\ell k},
\ee
where $m<n$. In other words, the matrix $M$ is the product of two rectangular matrices of size $n\times m$ and $m\times n$. Assume that $\rank B=m$. Otherwise,
if $\rank B=m'$ with $m'<m$, then the rows of $B$ are linearly dependent, and we can present $M$ as a sum of $m'$ terms:
\be{formM1}
M_{jk}=\sum_{\ell=1}^{m'} \tilde A_{j\ell}B_{\ell k}.
\ee

Obviously, $\det M=0$. Let us find zero eigenvectors of $M$. For this, we extend the matrix $B_{jk}$ by adding $n-m$ rows with elements $B_{m+1,k}, B_{m+2,k},\dots, B_{n,k}$. Let us
denote this extended $n\times n$ matrix by $\tilde B_{jk}$. We require that this extended matrix is invertible. Then zero eigenvectors $\Psi^{(a)}$,
$a=1,\dots,n-m$ have the following components
\be{Zev}
\Psi^{(a)}_j=(\tilde B^{-1})_{j,m+a}.
\ee

The proof is elementary:
\be{pro}
\sum_{k=1}^nM_{jk}\Psi^{(a)}_k= \sum_{k=1}^n\sum_{\ell=1}^m A_{j\ell}B_{\ell k} (\tilde B^{-1})_{k,m+a}
=\sum_{\ell=1}^m A_{j\ell}\delta_{\ell,m+a}=0,
\ee
because $\ell<m+a$.

Since we assume that $\tilde B_{jk}$ is invertible, the vectors $\Psi^{(a)}$ are linearly independent.

In particular, let $B$ be a rectangular Cauchy matrix
\be{BCau}
B_{jk}=\frac{\theta_1(x_j-y_k+\lambda)}{\theta_1(x_j-y_k)}, \qquad j=1,\dots,m, \qquad k=1,\dots n,
\ee
dependent on pairwise distinct complex numbers $x_{1},\dots, x_{m}$ and $y_{1},\dots, y_{n}$.
We can create an extended matrix as follows:
\be{BCau1}
\tilde B_{jk}=\frac{\theta_1(x_j-y_k+\lambda)}{\theta_1(x_j-y_k)},\qquad j,k=1,\dots n.
\ee
Here $x_{m+1},\dots, x_{n}$ are generic complex numbers not equal to $x_1,\dots,x_m$ and $y_1,\dots,y_n$.
Then the extended matrix $\tilde B_{jk}$ is non-degenerated, and the inverse matrix is
\be{BCau-inv}
(\tilde B^{-1})_{jk}=\frac1{\theta_1(\lambda)\theta_1(S+\lambda)}\frac{\theta_1(S+\lambda-x_k+y_j)}{\theta_1(x_k-y_j)}
\frac{\theta_1(x_k-\bar y)\theta_1(\bar x-y_j)}{\theta_1(x_k-\bar x_k)\theta_1(\bar y_j- y_j)},
\ee
where
$$
S=\sum_{j=1}^{n}(x_j-y_j).
$$
Respectively, zero eigenvectors of $M$ are
\be{ZevCa}
\Psi^{(l)}_j=C_l\theta_1(S+\lambda-x_{m+l}+y_j)
\frac{\theta_1(\bar x_{m+l}-y_j)}{\theta_1(\bar y_j- y_j)},
\ee
where $C_l$ are some constants. We see that different eigenvectors are parameterized by the numbers $x_{m+1},\dots, x_{n}$.

\section{Contour integral method\label{A-CIM}}

A contour integral method allows one to calculate or transform sums of a special form containing the Jacobi theta functions. To illustrate this method, we
consider two examples. The first example is related to the transformation of the matrix product $\mathbf{\Omega}^1\mathbf{\Omega}^0$.

It follows from \eqref{Omell} that the product $\mathbf{\Omega}^1\mathbf{\Omega}^0$ has the following form:
\be{OmOm-1}
\left(\mathbf{\Omega}^1\mathbf{\Omega}^0\right)_{jk}
%
=\frac{\theta_2(0)f(u_k,\bu_k)}{\theta_1(x)\theta_2(x)f(u_j,\bv)}H_{jk},
\ee
where
\be{HHH}
H_{jk}=\sum_{a=1}^{n-2p+1}  \frac{\theta_2(u_a-\bu)\theta_1(u_a-\bv)}{\theta_1(u_a-\bu_a)\theta_2(u_a-\bv)}
\frac{\theta_2(u_{\ell j}-x)\theta_1(u_{\ell k}+x)}
{\theta_2(u_a-u_j)\theta_2(u_a-u_k)}.
\ee
Consider a contour integral
\be{cI-2}
J=\frac{\theta'_1(0)}{2\pi i}\oint \,dz \frac{\theta_2(z-\bu)\theta_1(z-\bv)}{\theta_1(z-\bu)\theta_2(z-\bv)} \frac{\theta_2(z-u_{j}-x)\theta_1(z-u_{k}+x)}
{\theta_2(z-u_j)\theta_2(z-u_k)}.
\ee
Using \eqref{A-shift} we obtain that $J=0$ due to periodicity (recall that $\#\bu=n-2p+1$ and $\#\bv=n$). On the other hand, this integral equals the sum of the residues within the integration contour. The sum of the residues in $z=u_a$ gives $H_{jk}$. There are also poles in $z=v_q+1/2$, $q=1,\dots,n$.
Finally, there is a pole at $z=u_j+1/2$ for $j=k$. Thus, we obtain
\be{HH-res}
H_{jk}=\delta_{jk}\frac{f(u_j,\bv)}{f(u_j,\bu_j)}\frac{\theta_1(x)\theta_2(x)}{\theta_2(0)}
+\theta_2(0)\sum_{q=1}^{n}\frac{f(v_q,\bv_q)}{f(v_q,\bu)}
\frac{\theta_1(u_{j}-v_q+x)\theta_2(u_{k}-v_q-x)}
{\theta_1(u_j-v_q)\theta_1(u_k-v_q)}.
\ee
This implies
\begin{equation}\label{OmOm-2}
\left(\mathbf{\Omega}^1\mathbf{\Omega}^0\right)_{jk}=\delta_{jk}+
\frac{\theta_2^2(0)f(u_k,\bu_k)}{\theta_1(x)\theta_2(x)f(u_j,\bv)}\sum_{q=1}^{n}\frac{f(v_q,\bv_q)}{f(v_q,\bu)}
\frac{\theta_1(u_{j}-v_q+x)\theta_2(u_{k}-v_q-x)}
{\theta_1(u_j-v_q)\theta_1(u_k-v_q)}.
\end{equation}
Hence, we arrive at \eqref{OmOm-3}.

In the second example, we calculate the coefficients $\mathcal{A}_j$ \eqref{calAB}. We have
\be{calAj}
\mathcal{A}_j=\frac{-1}{f(v_n,v_j)\theta_1(x)\theta_1(x-S)}\frac{\theta_2(\bu-v_j)}{\theta_2(\bv_{n,j}- v_j)}
\frac{f(v_n,\bv_n)}{f(v_n,\bu)}G^a_j,
\ee
where
\be{gaj}
G^a_j=\sum_{a=1}^{n-1} \frac{\theta_1(u_a-v_j-x+S )\theta_1(u_a-v_n+x)}{\theta_1(u_a-v_j)\theta_1(u_a-v_n)}
\frac{\theta_1(u_a-\bv_n)}{\theta_1(u_a-\bu_a)}.
\ee
The function $G^a_j$ can be found from the following contour integral:
\be{Jaj}
J^a_j=\frac{\theta'_1(0)}{2\pi i}\oint\; \dd z \frac{\theta_1(z-v_j-x+S)\theta_1(z-v_n+x)}{\theta_1(z-v_j)\theta_1(z-v_n)}
\frac{\theta_1(z-\bv_n)}{\theta_1(z-\bu)}.
\ee
Due to the periodicity, this integral is zero. On the other hand, by virtue of the residue theorem, it gives $G^a_j$ and residue at the pole $z=v_n$
\be{Jaj1}
J^a_j=0= G^a_j +\theta_1(x) \frac{\theta_1(v_{nj}-x+S)}{\theta_1(v_{nj})}
\frac{\theta_1(v_n-\bv_n)}{\theta_1(v_n-\bu)},
\ee
leading to \eqref{calAj-res}.


\end{document}